# Folding model calculations for $^6$He+$^{12}$C Elastic Scattering

**Awad A. Ibraheem**[1,2*]

1. Physics Department, King Khalid University, Abha, Saudi Arabia.
2. Physics Department, Al-Azhar University, Assiut 71524, Egypt.

## ABSTRACT

In the framework of the double folding model, we used the α+2n and di-triton configurations for the nuclear matter density of the $^6$He nucleus to generate the real part of the optical potential for the system $^6$He+$^{12}$C. As an alternative, we also use the high energy approximation to generate the optical potential for the same system. The derived potentials are employed to analyze the elastic scattering differential cross section at energies of 38.3, 41.6 and 82.3 MeV/nucleon. For the imaginary part of the potential we adopt the squared Woods-Saxon form. The obtained results are compared with the corresponding measured data as well as with available results in literature. The calculated total reaction cross sections are investigated and compared with the optical limit Glauber model description.

## I. INTRODUCTION

During the last three decades, the formation of nuclear halos has become one of the most interesting phenomena in the nuclear landscape. The two-neutron halo nucleus $^6$He, as a prototype of the Borromean structure, has attracted the attention of most experimental and theoretical physicists [1-16]. Elastic scattering of $^6$He from a target nucleus is considered as a gate way reaction in order to investigate characteristics of weakly bound light nuclei and help us understand their structure [10]. The elastic scattering data at 38.3 and 41.6 MeV/nucleon from $^{12}$C have been analyzed within the framework of the optical model [8, 17]. Initial analyses were done using a microscopic real potential based on different versions of the density-dependent M3Y(DDM3Y) effective interaction and the usual phenomenological Woods-Saxon (WS)

---

*awad_ah_eb@hotmail.com

imaginary potential. From these analyses, it was concluded that a satisfactory description of the whole angular range of the data could not be obtained by adjusting the imaginary potential parameters or using a simple renormalization of the real potential [5]. A systematic analysis of the $^6$Li scattering data at similar energies revealed that a good description of the data could be obtained when a repulsive empirical dynamic polarized potential (DPP) of a surface form was added to the real part of the potential and an absorptive surface form to the imaginary part [4, 17]. Another analysis of $^6$He and $^6$Li elastic scattering at about 35 MeV/nucleon showed that the data could be successfully described by optical model potentials with relatively deeper imaginary potential for $^6$Li [3].Lukyanov*et al*. used microscopic direct and exchange real part as well as microscopic imaginary part based on the high-energy approximation (HEA)[18] with a minimal number of free parameters to study $^6$He+$^{12}$C elastic scattering.

The main objective of the present work is two-fold:-First, we want to explore the possibility of adopting another structure of the $^6$He nucleus within the Triton + Triton (t+t) approximation besides the three-particle approximation (α+2n). In a recent work, Giot*et al*[4,19] used a *2n* transfer reaction and found that the spectroscopic factor for the α+2n configuration is on the expected line, while in the case of t+t configuration, it is much smaller than the theoretical description. The second objective is to check the ability of the imaginary part of the squared Woods Saxon (WS2) form instead of the usual WS shape to describe the experimental data of $^6$H+ $^{12}$C elastic scattering. However, the need for a renormalization of the $^6$He+$^{12}$C potential has been suggested by several authors [17,18]. Lapoux *et.al* [17] performed folding analysis for the$^6$He + $^{12}$C system and suggested the need for an appropriate DPP to be added to the folded potential. Recently, Aygun *et.al* [20] proposed a new parameterization for the real and imaginary parts of the DPP with the sum of the standard WS and derivative WS form factors. At intermediate energies, the DPP is considered to arise mainly from the strong coupling to the breakup channels of $^6$He. This has been clearly demonstrated in the four-body CDCC analysis done by the Kyushu group [21].They found that the polarization potential due to breakup channels is repulsive, which ultimately leads to the value of the normalization constant being less than unity in the double folding (DF) analysis.

In view of the above discussion, we re-examined the elastic scattering of $^6$He from $^{12}$C target at three different energies, 38.3, 41.6 and 82.3 MeV/nucleon, using several density distributions of $^6$He nucleus. We performed the calculations using the microscopic double folding (DF) model. In the folding calculation we used densities based on $\alpha$ +2n and t+t models (denoted as D1 and D2, respectively) as well as with the few body Faddeev calculation (denoted as Q3) for the density of the $^6$He nucleus [6,16].

In this work, we have successfully found that the $^6$He+$^{12}$C data [1, 2, 8, 17] can be reproduced and interpreted within the framework of a DF real potential supplemented by an imaginary part of WS2 form.

## FORMALISM

### II-1 Theoretical potential within the DF model

The optical model potential involved in this work has the standard form,

$$U(R) = V_C(R) + V_r(R) + i\, W(R), \tag{1}$$

where $V_r(R)$ and $W(R)$ are the attractive real and imaginary parts of the nuclear potential, respectively.

$V_C(R)$ is the Coulomb potential due to a uniform distribution of appropriate size, and radius $R_c = 1.13(A_P^{1/3} + A_T^{1/3})$, so then

$$V_C(R) = \begin{cases} \frac{Z_P Z_T e^2}{R}, & R > R_c \\ \frac{Z_P Z_T e^2}{2R_c}\left[3 - \left(\frac{R}{R_c}\right)^2\right], & R \leq R_c \end{cases} \tag{2}$$

$A_P$ and $A_T$ are the mass numbers of the projectile ($P$) and the target ($T$) nuclei, while $Z_P$ and $Z_T$ denote their corresponding charges, respectively. In the phenomenological analysis the attractive real and imaginary potentials are treated phenomenologically using conventional forms like WS potentials or any other form. Alternative analyses replace the phenomenological real part of equation (1) by a microscopic one based on the DF approach. The DF potential may be written as the double-convolution integral

$$V_r(R) = N_r \int \rho_p(\vec{r_1})\rho_T(\vec{r_2}) v_{NN}(|\vec{s}|) d^3 r_1 d^3 r_2, \vec{s} = \vec{R} - \vec{r_1} + \vec{r_2} \tag{3}$$

where $\rho_p$ and $\rho_T$ are the ground state density distributions of projectile ($^6$He) and target ($^{12}$C) nuclei as well as the renormalization factor $N_r$. The effective nucleon–nucleon (NN) interaction, $v_{NN}$, is integrated over both density distributions. Several NN interaction expressions can be used for the folding model potentials. Among the different kinds of effective interactions, the so-called M3Y (Michigan 3 Yukawa) interaction was widely used in many folding calculations of the heavy ion HI optical potential [22]. In the present work, we use this form of interaction with the relevant exchange correction term due to the Pauli principle, given by

$$v_{eff}^{M3Y}(|\vec{s}|, E) = v_{D}^{M3Y}(|\vec{s}|) + v_{EX}^{M3Y}(|\vec{s}|, E) \tag{4}$$

where $v_{D}^{M3Y}(|\vec{s}|)$ is the direct term of the NN effective interaction and $v_{EX}^{M3Y}(|\vec{s}|, E)$ is the corresponding knock on exchange term. Based on equation (4), the DF potential is expressed as

$$V_r(E, R) = V_D(R) + V_{EX}(E, R) \tag{5}$$

where $V_D(R)$ is the direct part and $V_{EX}(E, R)$ is the exchange part of the real folded potential. Since there are few versions of the M3Y effective NN interaction it is worth mentioning that we use the Reid form. The explicit form of the Reid effective interaction is

$$v_{D}^{M3Y}(|\vec{s}|) = 7999 \frac{e^{-4s}}{4s} - 2134 \frac{e^{-2.5s}}{2.5s} \tag{6}$$

For the exchange part we have two different choices. The first is the zero range one, given as

$$v_{EX}^{M3Y}(|\vec{s}|, E) = -276 (1 - 0.005E)\delta(s) \tag{7}$$

The second is the finite range knock-on exchange contribution [18], and is computed from the following relation:

$$V_{EX}(E, R) = \int \rho_p(\vec{r_1}, \vec{r_1} + \vec{s})\rho_T(\vec{r_2}, \vec{r_2} - \vec{s})v_{EX}(\rho, E, s)\exp[\frac{i\vec{K}(E,R)\vec{s}}{M}]d^3r_1 d^3r_2 \tag{8}$$

This exchange term is nonlocal. However, an accurate local approximation can be obtained by treating the relative motion locally as a plane wave [32, 33]. So, the local momentum of relative motion $K(E, R)$ can expressed as

$$\left|\vec{K}(E, R)\right|^2 = \frac{2\mu}{\hbar^2}[E_{CM} + V_{DF}(E, R) - V_C(R)] \tag{9}$$

where μ is the reduced mass, $\mu = \frac{A_P A_T}{A_P + A_T}$ and $E_{CM}$ is the relative energy in the center-of-mass system.

## II.2 Optical potential within the HEA

It is of interest to carry out a similar analysis on a fully microscopic basis, where both real and imaginary parts of the optical potential can be calculated. In this section, we intend to test the optical potential using the so-called HEA aiming to study the effects of the different behaviors of $^6$He densities to explain the available data of $^6$He+$^{12}$C differential cross-sections at 38.3, 41.6 and 82.3 MeV/N. First, we can calculate the elastic cross-sections of $^6$He+$^{12}$C by using two free parameters ($N_r$ and $N_I$) which renormalize the depths of the real ($V^H$) and imaginary ($W^H$) parts of the HEA optical potentials in the form [18]

$$U^{Opt} = N_r V^H + i N_I W^H \tag{10}$$

where

$$V^H = -\frac{\hbar v}{(2\pi)^2} \bar{\sigma}_{NN} \bar{\alpha}_{NN} \times \int_0^\infty \rho_1(q)\rho_2(q) f_N(q) j_0(qr) q^2 dq \tag{11}$$

$$W^H = -\frac{\hbar v}{(2\pi)^2} \bar{\sigma}_{NN} \times \int_0^\infty \rho_1(q)\rho_2(q) f_N(q) j_0(qr) q^2 dq \tag{12}$$

with $v$ being the velocity of the nucleus-nucleus relative motion, $\rho(q)$ being the form factors corresponding to the nucleon density distributions of the nuclei and $f_N(q)$ being the amplitude of the NN scattering which depends on the transfer momentum q and $\bar{\alpha}_{NN}$ are averaged over the isospin of the nucleus total NN scattering cross section and the ratio of the real to imaginary parts of the forward NN scattering amplitude, respectively.

## II.3 Nuclear density of $^6$He

There are some uncertainties concerning the density of $^6$He. So, several choices have been adopted to study the effect caused by the halo structure of $^6$He [6-18]. For this reason, three different forms of the ground state density distribution are used in the present folding calculation. In the first form, the $^6$He nucleus is assumed to consist of a core of $^4$He and two halo neutrons ($\alpha$+2n). Then, one may formulate the nuclear matter density of $^6$He as

$$\rho_{^6He}(r) = \int |\psi(R')|^2 \left(\rho_c\left(\left|\bar{r}-\frac{1}{3}\overline{R'}\right|\right) + \rho_{2n}\left(\left|\bar{r}+\frac{2}{3}\overline{R'}\right|\right)\right) d\overline{R'} \tag{13}$$

where the core-halo relative wave function is represented in the form

$$\psi(R') = \frac{4\alpha}{\sqrt{15}}\left(\frac{2\alpha}{\pi}\right)^{3/4} R'^2 \exp(-\alpha R'^2) \ . \tag{14}$$

The parameter α= 0.341 fm$^{-2}$ is adjusted to reproduce the experimental value for the root mean square (rms) radius of $^6$He= 2.54 fm [6]. This radius is close to that evaluated from the four-body analysis of $^6$He+$^{12}$C total reaction cross sections [6] as well as by the analysis of elastic scattering of $^6$He on protons at high energies [7]. The core ($^4$He) and halo densities are taken in a Gaussian form, respectively, as [19]

$$\rho_c(r) = 4(\frac{\gamma_c}{\pi})^{3/2} \exp(-\gamma_c r^2) \tag{15}$$

and

$$\rho_{2n}(r) = 2(\frac{\gamma_h}{\pi})^{3/2} \exp(-\gamma_h r^2) \tag{16}$$

where $\gamma_c$ = 0.6756 fm$^{-2}$ and $\gamma_h$= 0.1305 fm$^{-2}$. The resulting density, denoted as D1, yields rms radii of 1.49 and 3.39 fm for the free $^4$He and two-neutron, respectively.

Similar to the $\alpha$ +2n cluster model, we also introduced the nuclear matter density of $^6$He based on the Triton-Triton (t-t) cluster model as

$$\rho_{^6He}(r) = \int |\psi(R)|^2 (\rho_{^3H_1}(|\vec{r} - \frac{\vec{R}}{2}|) + \rho_{^3H_1}(|\vec{r} + \frac{\vec{R}}{2}|)) d\vec{R} \tag{17}$$

where R is the $^3$H$_1$-$^3$H$_1$ separation inside the $^6$He nucleus, and $\psi(R)$ is the wave function of the relative motion of $^3$H$_1$-$^3$H$_1$ clusters in the ground state of the $^6$He nucleus. This relative wave function is taken in the same form as equation (14) with the parameter $\alpha = 0.11495$ fm$^{-2}$ to obtain the same value of the rms radius of $^6$He (2.54 fm). A Gaussian form for the density distribution of the triton ($^3$H$_1$) has been assumed as [23]

$$\rho_{^3H_1}(r) = 3(\frac{\gamma_1}{\pi})^{3/2} \exp(-\gamma_1 r^2) \tag{18}$$

with $\gamma_1 = 0.5577\ fm^{-2}$. The calculations have been performed with the code MATHEMATICA [24]. The resulting density is denoted by D2. Finally, for the sake of comparison, we considered another form for $^6$He density which is taken from Refs. [6,7] obtained by Faddeev wave function and denoted by Q3. Throughout the present work, this density Q3 can be expressed as a summation of thirteen Gaussian terms as

$$\rho_{^6He}(r) = \sum_{k=1}^{13} c_k \exp\left[-a_k r^2\right] \text{fm}^{-3} \tag{19}$$

The parameters $c_k$ and $a_k$ are listed in Table 1. It should be noted that the first term represents the matter density distribution of alpha particle with rms radius equal to 1.47 fm. The obtained densities are shown in Fig.1. It is noticed that the D1 density seems substantially deeper than D2 and Q3 at the center (r = 0.0fm); however, the D1 and Q3 densities have consistent values through the radial range r = 3.0- 5.0 fm. One may notice that all densities have the same value (~0.056 fm$^{-3}$) at r $\cong$1.78 fm. For comparison, we plot in Fig. 1 the matter density of $^6$He obtained from realistic microscopic calculations with the large-scale shell-model (LSSM) method [25].The ground state matter density of $^{12}$C is taken as a two parameter Fermi function as [26]

$$\rho(r) = \frac{0.207}{1+exp(\frac{r-2.1545}{0.425})} \quad \text{fm}^{-3} \quad (20)$$

This density yields a rms radius equal to 2.298 fm, close to those obtained from (*e,e*) scattering measurements. This density has a similar shape to that obtained by shell model calculations [26].

**Table 1:** The parameters $c_k$ and $a_k$ obtained from Eq. (20).

| k | $c_k$ (fm$^{-3}$) | $a_k$ (fm$^{-2}$) |
|---|---|---|
| 1 | 0.4154230 | 0.694153 |
| 2 | -2.208400 | 0.603806 |
| 3 | 0.0018019 | 0.063792 |
| 4 | 0.5811340 | 0.522915 |
| 5 | 0.5696400 | 0.522679 |
| 6 | -2.380550 | 0.424470 |
| 7 | 0.5809250 | 0.522914 |
| 8 | 0.9387690 | 0.355672 |
| 9 | 0.5797400 | 0.522909 |
| 10 | 0.5678100 | 0.523446 |
| 11 | 0.5757130 | 0.522877 |
| 12 | -0.608940 | 5.640160 |
| 13 | 0.6015380 | 5.509230 |

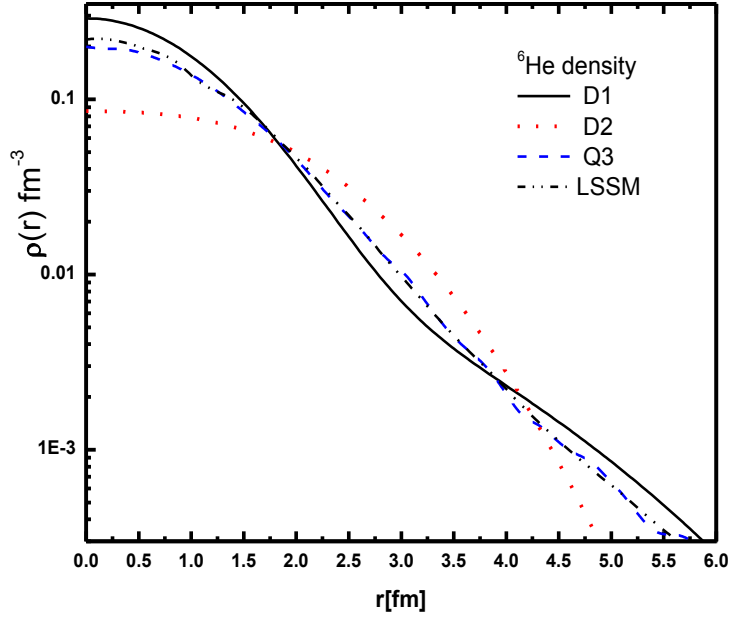

**FIG. 1.** The nuclear matter density distribution of $^6$He (D1 and D2) deduced from Eqs.(13) and (17) as compared with the Faddeev model (Q3) obtained in Eq.(19) and with the LSSM model.

## III.   PROCEDURE

The analysis was done using a real part of the optical potential obtained microscopically by the DF model of equation (3). In this model, we used three different forms of the nuclear matter density distribution of $^6$He (D1, D2 and Q3) folded with a realistic M3Y effective NN interaction. The resulted potentials with each density are denoted as DFC1, DFC2 and DFC3, respectively. This analysis was carried out using the HIOPTIM-94 program [27], which was fed with the calculated microscopic real potentials supplemented by the imaginary part of the squared Woods-Saxon (WS2) form,

$$W(R) = \frac{W_0}{\left(1 + \exp\left(\frac{R - R_I}{a_I}\right)\right)^2}, \quad R_I = r_I \left(A_P^{1/3} + A_T^{1/3}\right) \tag{21}$$

where $W_0$, $r_I$ and $a_I$ are the depth, radius and diffuseness parameters, respectively. Best fits were obtained by minimizing the $\chi^2$ value, where

$$\chi^2 = \frac{1}{N} \sum_{k=1}^{N} \left[ \frac{\sigma_{th}(\theta_k) - \sigma_{exp}(\theta_k)}{\Delta\sigma_{exp}(\theta_k)} \right]^2. \tag{22}$$

$\sigma_{th}$ ($\sigma_{exp}$) is the theoretical (experimental) cross section at angle $\theta_k$ in the center-of-mass system, $\Delta\sigma_{exp}$ is the experimental error and $N$ is the number of data points. The $^6$He+$^{12}$C elastic scattering differential cross sections were calculated at energies 38.3, 41.6 and 82.3 MeV/nucleon. The search was carried out on three free parameters: the imaginary WS2 potential parameters, $W_0$, $r_I$ and $a_I$. The renormalization factor $N_r$ for the real part of the DF potential was fixed at unity.

## IV. RESULTS AND DISCUSSION

In this work, we investigated the ability of the derived DF potentials based on α+2n or the di-triton configuration to analyze the elastic scattering of $^6$He + $^{12}$C at 38.3, 41.6 and 82.3 MeV/nucleon. The best fit parameters along with the corresponding minimum $\chi^2$ values for the differential cross-section and reaction cross-section $\sigma_R$ are listed in Table (2). In earlier studies, folding analyses performed for the $^6$He + $^{12}$C system clearly indicated that the data require a strong reducing factor Nr of the real part to be correctly described with the standard imaginary WS potential. In our calculations, we replaced the standard WS form by the WS2 form in Eq. (21). The angular distributions of the differential cross section at the above-mentioned energies were calculated using the microscopic real potentials (DFC1, DFC2 and DFC3) with the best fit parameters as shown in Table (2) using the HIOPTM-94 code. The results are compared with the corresponding measured data as shown in Fig. (2). From this figure it is noticed that for the energies 38.3 and 41.6 MeV/nucleon, apparently, a good description of the elastic scattering is obtained using the DFC1, DFC2 and DFC3 potentials based upon the M3Y interaction. The only exception is the largest measured scattering angles around $20^0$. The same behavior was also noticed when these data were previously analyzed using other folded potentials [8,17]. For the energy 82.3 MeV/nucleon, all the calculated potentials in the present work reveal similar descriptions except at the forward angles $<5^0$.

The reaction (absorption) cross section, $\sigma_R$, is considered an important quantity in the analysis of elastic scattering reactions. Hence, it would be interesting to investigate whether one can deduce a reasonable determination of $\sigma_R$ using the derived potentials. In this context, we plot the obtained $\sigma_R$ values using the three folded potentials versus energy as shown in the lower part of Fig. 3. As shown in Table (2), a strong energy dependence is found where $\sigma_R$ decreases with increasing energy. This reflects the behavior of probability of absorption with increasing energy. It is interesting to notice that the obtained value of $\sigma_R$ at 229.8 MeV agrees well with that found from the analysis of $^6Li+^{12}C$ elastic scattering at 210 MeV [28]. It is worth mentioning that Lou *et al* [1,2] used two sets of the imaginary parts of optical potential to fit the experimental data at 82.3 MeV/nucleon including contributions from the inelastic channels with a real DF potential based on CDM3Y6 effective NN interaction[3]. These two sets give total reaction cross-sections $\sigma_R$ of 853 and 843 mb.

We also used the optical limit Glauber model approximation (OLA) to calculate the total reaction cross section and compared it with our model. Within the OLA the reaction cross section $\sigma_R$ is expressed as [29-31]:

$$\sigma_R = 2\pi \int_0^\infty b \, db \left(1 - T(b)\right) \tag{23}$$

where T (b) is the transparency function of the collision at impact parameter b defined as

$$T(b) = \exp\left(-\sigma_{NN} \int d^2 \vec{b_1} \rho_P^z(b_1) \rho_T^z(|\vec{b} - \vec{b_1}|)\right). \tag{24}$$

In Eq. (24) $\sigma_{NN}$ is the nucleon-nucleon cross section at the appropriate *NN* relative energy, with $\rho_i^z(b)$ the corresponding thickness functions. Further, extensive details of the average $\sigma_{NN}$ and $\alpha_{NN}$ in terms of proton number ($Z_P$ and $Z_T$) and neutron number ($N_P$ and $N_T$) of the projectile and target nuclei can be obtained, see Refs. [27,29, 30]. The lower panel of Fig. 3 shows the energy dependence of the total reaction cross section resulting from the optical limit phase shift calculations with D1, D2 and Q3, namely OLA(D1), OLA(D2) and OLA(Q3), respectively. In general, the obtained total reaction cross sections showed in this figure decreases linearly as energy increases. The obtained value of $\sigma_R$ compared to the corresponding one of the WS of Lou *et al* [1,2] is 20% larger than that at 82.3 MeV/nucleon.

The other important information is the volume integral $J_U$ which can be obtained from the elastic scattering. For an interaction potential U(R) between two nuclei, the volume integral per interacting nucleon pair $J_U$ can be defined as

$$J_U = \frac{1}{A_P A_T} \int U(R) d\vec{R} \tag{25}$$

This quantity is currently used as a sensitive measure of the potential strength. In the present work, we apply this definition to the real and imaginary parts of U(R), denoted as $J_R$ and $J_I$, respectively. Since we fix the normalization factor to unity for the real part we expect no energy dependence for the real part but we have this dependence for the imaginary one. The energy dependence for the imaginary volume integral is shown in Table2 and in the upper part of Fig.3. It is obvious that the obtained reaction cross section and the imaginary volume integral have almost identical energy dependence. This result is physically expected where both $J_I$ and $\sigma_R$ involve absorption to nonelastic channels.

**Table 2.** Parameters of the optical potential obtained for the $^6$He+$^{12}$C system at 38.3, 41.6 and 82.3 MeV/nucleon. The real folded potential is calculated with the M3Y interaction using D1 and D2 densities or the Q3 density. Both have rms radius equal to 2.54 fm. The normalization factor Nr is equal to unity. The parameters are the volume depth ($W_0$) in MeV, radius and diffuseness parameters($r_I$ and $a_I$) in fm, real and imaginary volume integrals ($J_R$ and $J_I$) in MeV.fm$^3$, real and imaginary rms radii ($R_R$ and $R_I$ respectively), total reaction cross section ($\sigma_R$) in mb and the best fit $\chi 2$.

| Pot | $N_r$ | $W_0$ (MeV) | $r_I$ (fm) | $a_I$ (fm) | $J_R$ (MeVfm$^3$) | $J_I$ (MeVfm$^3$) | $\chi 2$ | $\sigma_R$ (mb) | $R_R$ (fm) | $R_I$ (fm) |
|---|---|---|---|---|---|---|---|---|---|---|
| colspan E=82.3 MeV/nucleon ||||||||||| 
| DFC1 | 1.0 | 95.82 | 1.0261 | 0.8830 | 308.20 | 296.07 | 69.30 | 878.3 | 3.89 | 3.53 |
| DFC2 | 1.0 | 89.14 | 1.0395 | 0.8830 | 308.20 | 288.52 | 70.50 | 845.5 | 3.89 | 3.56 |
| DFC3 | 1.0 | 83.82 | 1.0797 | 0.8830 | 308.20 | 304.60 | 84.10 | 930.7 | 3.89 | 3.67 |
| E=41.6 MeV/nucleon |||||||||||
| DFC1 | 1.0 | 26.323 | 1.2908 | 0.8830 | 368.97 | 166.20 | 14.60 | 1078.0 | 3.82 | 4.22 |
| DFC2 | 1.0 | 33.795 | 1.2358 | 0.8830 | 368.97 | 186.33 | 8.40 | 1064.0 | 3.82 | 4.07 |
| DFC3 | 1.0 | 20.176 | 1.3231 | 0.8830 | 368.97 | 137.60 | 12.70 | 1058.0 | 3.82 | 4.31 |
| E=38.3 MeV/nucleon |||||||||||
| DFC1 | 1.0 | 13.460 | 1.4379 | 0.8830 | 363.58 | 119.21 | 4.83 | 1117.0 | 3.83 | 4.62 |
| DFC2 | 1.0 | 17.362 | 1.3776 | 0.8830 | 363.58 | 134.43 | 6.87 | 1109.0 | 3.83 | 4.45 |
| DFC3 | 1.0 | 10.84 | 1.4785 | 0.8830 | 363.58 | 101.32 | 5.2 | 1092.0 | 3.83 | 4.73 |

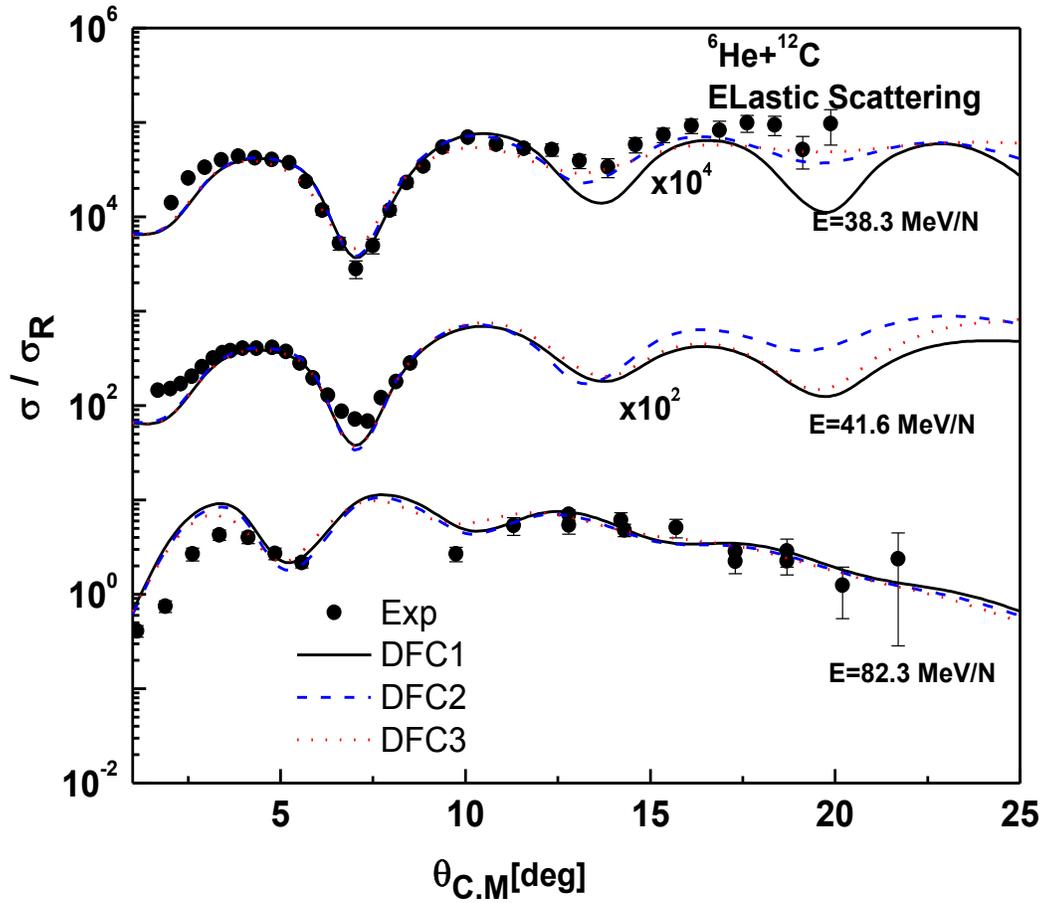

Fig 2. Elastic scattering data for $^6$He on $^{12}$C at 38.3, 41.6 and 82.3 MeV/nucleon in comparison with the results given by the real DFC1, DFC2 and DFC3 potentials obtained with the M3Y interaction. Experimental data are taken from Refs.[1, 2, 8, 17].

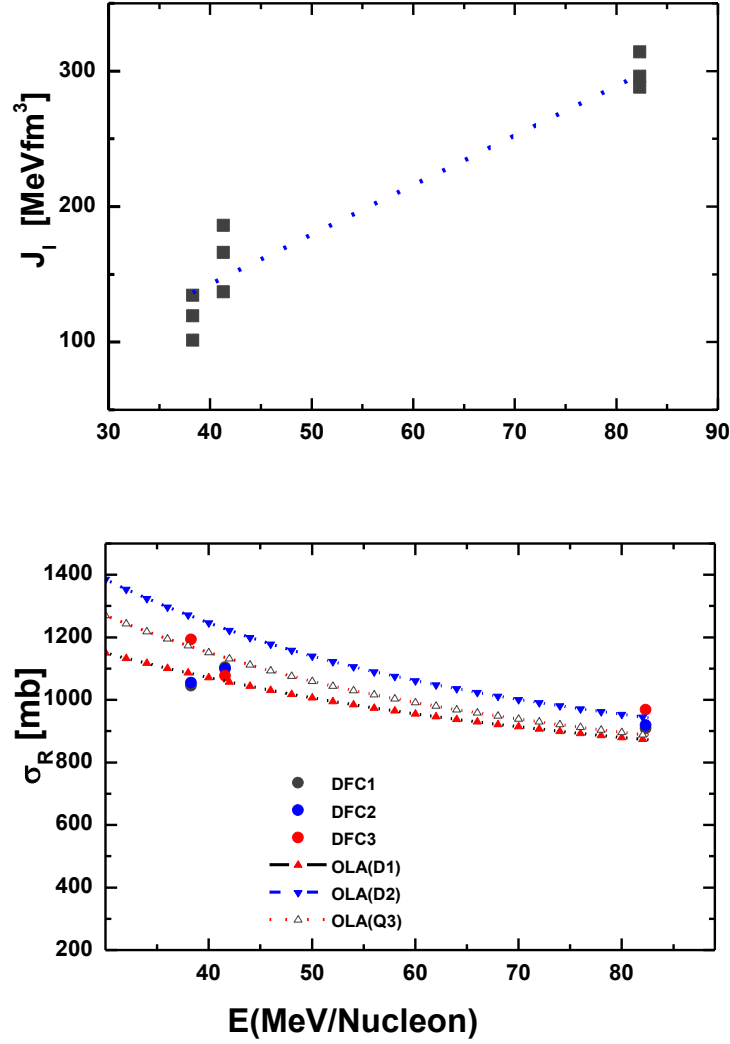

Fig 3. Energy dependence of the obtained imaginary volume integral and reaction cross section. The lines are drawn only to guide the eye.

To validate our analysis, we plotted expression (11) at 82.3 MeV/nucleon in comparison with the DF potential with zero-range (DF-D1Z, DF-D2Z,DF-Q3Z) and finite-range (DF-D1F,DF-D2F,DF-Q3F) exchange contribution for the considered densitiesD1,D2 and Q3, as shown in Fig.4. First we note from this figure that the zero–range potential is deeper than the finite range density by about 50% at the center ($R$=0), while they resemble each other for the radial distance R from 4 fm up to 7 fm. Second, we can observe that the different DF and HEA potentials are in agreement with each other in the vicinity of the strong absorption radius $R_S$.

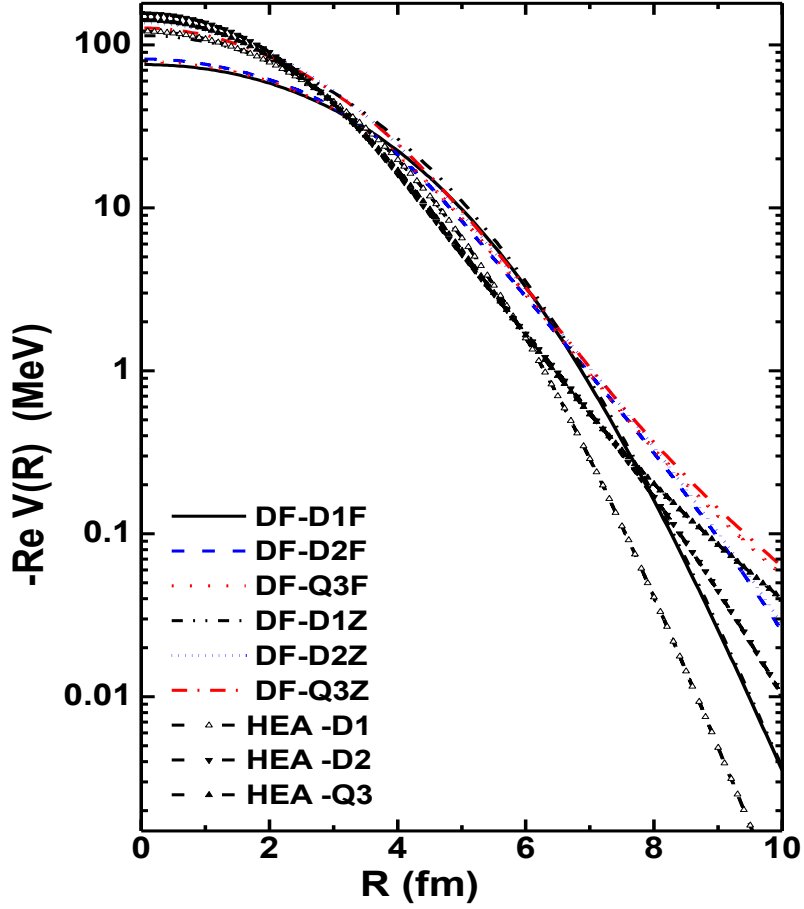

FIG. 4. Un-normalized $^6$He+$^{12}$C DF potentials extracted from expression (4, 11) for zero range and finite range at 82.3 MeV/nucleon compared with the HEA model.

Figure 5. presents the results of calculations of the $^6$He+$^{12}$C elastic cross-sections for energy E = 82.3 MeV/$N$ for all three densities (D1), (D2) and (Q3). One can see there is agreement with the data using $N_r$ = 0.85 , $N_I$ = 0.80 for the cases (D2) and (Q3) and using $N_r$ = 0.8, $N_I$ = 0.92 for case (D1). For the other two energies, there is agreement with the data using $N_r$ = 0.85 , $N_I$ = 0.45 for the cases (D1) and (D2) and using $N_r$ = 1.0, $N_I$ = 0.55 for case (Q3). Hence, we can conclude that for both cases, renormalization is necessary. The values

of $N_r$ and $N_I$ were chosen starting from the values $N_R = 1$ and $N_I = 1$ and decreased gradually in order to achieve a reasonable fit with the experimental data.

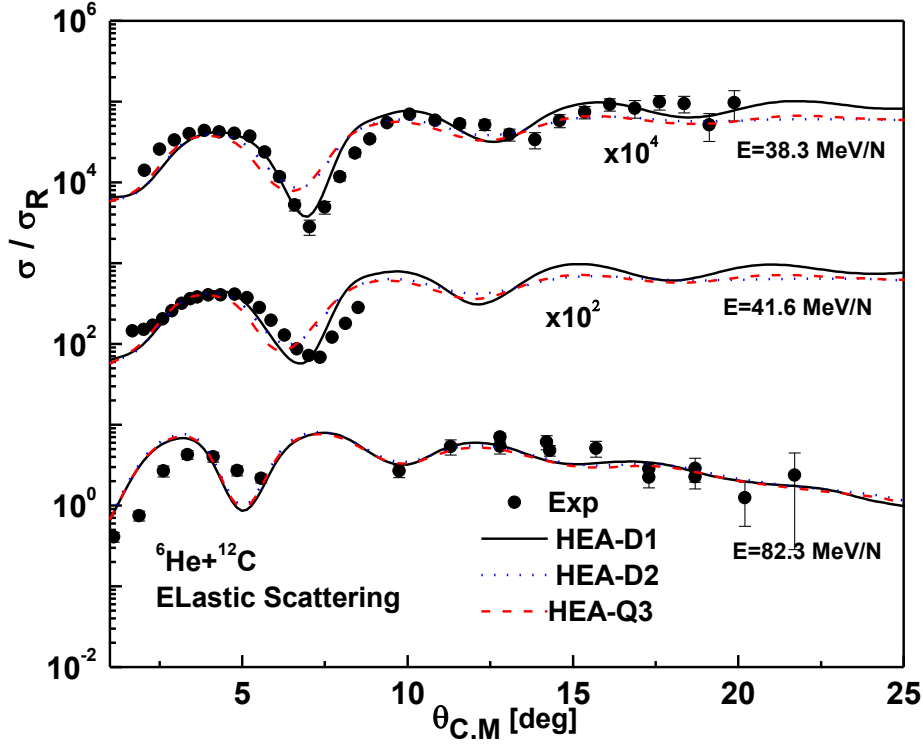

**Fig.5**: Elastic $^6$He+ $^{12}$C scattering cross sections at different energies calculated using $U^{Opt} = N_r V^H + iN_I W^H$ for various values of the renormalization parameters $N_r$ and $N_I$, giving a reasonable agreement with the data (presented in the text). Experimental data are taken from Ref.[1, 2, 8, 17]

## V. CONCLUSION

This work presents an optical model analysis of the $^6$He+$^{12}$C reactions at three different energies using microscopic DF optical model potentials. Several prescriptions for the $^6$He matter density (D1, D2 and Q3) have been used and compared with referenced data to understand the effects of these densities on the elastic scattering observables. The DF real potentials based upon the M3Y interaction have been successful in reproducing the scattering data over the measured angular ranges without renormalization factors $N_r$. It has been observed that these density distributions (D1, D2 and Q3) produce similar potentials and hence the elastic scattering angular distributions obtained by these potentials show similar behaviors. However, all densities used in the potential calculations provide an apparently good description for the elastic scattering cross section data of this system. This raises the question of the partial contribution of each suggested configuration into the whole $^6$He wave function and their respective cross sections. The imaginary part is parameterized during the analysis in a WS2 form instead of the usual WS shape. This gives an interesting result, which is in contrast with earlier work where contributions from inelastic excitations of the carbon target were used in a microscopic JLM potential with a reduction of the imaginary part. On the other hand, in Ref. [17] the renormalized DF real potential based upon the density-dependent CDM3Y6 effective NN interaction supplemented by the usual WS imaginary potential produced a satisfactory description of the 38.6 MeV/nucleon elastic scattering data. In the present calculation the unnormalized (Nr=1.0) DF potential based on the density-independent M3Y effective interaction could not successfully reproduce the same set of data using the normal WS imaginary potential. However, when the WS form is replaced by the WS2 form for the imaginary potential, a satisfactory description is obtained. It should be noted also that the forward angle (between 2°- 5°) data at 82.3 MeV/nucleon has not been described in a quantitative way.

For the sake of comparison, the application of the microscopic optical potentials with an imaginary part obtained within the HEA is justified in calculations of elastic scattering cross sections of the $^6$He+$^{12}$C data at different energies using only two (real and imaginary) free parameters to renormalize the depths of the real and imaginary parts. The best agreements with the data have been achieved when the D1 density of $^6$He is used. Therefore, the HEA model is much better than the DF model based upon the M3Y effective NN interaction for studying the nuclear structure effect of the $^6$He+$^{12}$C elastic scattering data at the three considered energies. Further studies on similar systems are recommended to take into account the dynamic polarization potential and inelastic scattering considerations.

## ACKNOWLEDGEMENT

I would like to thank Prof. Dr. M. El-Azab Farid, Dr. M. Ajmal Khan and Dr. Z. M. Mahmoud for reviewing the manuscript and hence their valuable comments.

# References

[1] J.L.Lou, Y.L.Ye, D.Y.Pang *et al.,* Phys. ReV.**C83**(2011) 034612.

[2] J.L.Lou, Y.L.Ye, D.Y.Pang *etal., Proceedings of the 14th National Conference on Nuclear Structure in China* (2012) pp. 158.

[3] Dao T. Khoa, Phys. ReV.**C 63** (2001) 034007.

[4] L. Giot,*et.al* ; Phys. ReV.**C71**(2005) 064311

[5] N. Keeley, K.W. Kemper, O. Momotyuk and K. Rusek, Phys. ReV**. C 77** (2008) 057601.;N. Keeley, N. Alamanos, K.W. Kemper and K. Rusek, Progress in Particle and Nuclear Physics 63 (2009) 396.

[6] J.S. Al-Khalili, J.A. Tostevin, and I.J. Thompson, Phys. ReV.**C54**(1996)1843 .

[7] J.S. Al-Khalili and J. A. Tostevin, Phys. ReV.**C57**(1998)1846.

[8] J. S. Al-Khalili, Phys. Lett.B**378**(1996)45 .

[9] Z. M. M. Mahmoud, Awad A. Ibraheem and M. El-Azab Farid, Int. J. Mod. Phys**E 23** (2014)1450008 .

[10] Z. M. M. Mahmoud, Awad A. Ibraheem and S. R. Mokhtar, Int. J. Mod. Phys**E 22** (2013) 1350086.

[11] W. R. Alharbi, Awad A. Ibraheem and M. El-Azab Farid, Journal of the Korean Physical Society **63** (2013)965.

[12] Awad A. Ibraheem, *et.al,* Physics of Atomic Nuclei **75** (2012)969.

[13] M Aygun, Y Kucuk, I Boztosun and Awad A. Ibraheem, Nucl. Phys. **A 848** (2010)245.

[14] M. El-Azab Farid, *et al.,* Arab Journal of Nuclear Sciences and Applications **43** (2010)169.

[15] M El-Azab Farid, A. M. A. Nossair and Awad A. Ibraheem, Int. J. Mod. Phys**E17** (2008)715

[16] M. V. Zhukov *et al.,*Nucl. Phys**. A 552** (1993), 353.

[17] V. Lapoux*et al* , Phys. ReV.**C66** (2002) 034608.

[18] V. K. Lukyanov*et al* , Phys. ReV.**C82** (2010) 024604.

[19] Xin-Shuai Yan, Jian-Song Wang, and Yan-Yun Yang, Chinese Physics **C 35**(2011) 550 .

[20] M. Aygun, I. Boztosun and K. Rusek; Modern Physics Letters **A28**(2013) 1350112

[21] T. Matsumoto *et al* , Phys. ReV.**C** 70 (2004)061601.

[22] G. R. Satchler and W. G. Love, Phys. Rep. **55** (1977)183.

[23] J. Cook and R. J. Griffiths, Nucl. Phys. **A366** (1981) 27.

[24] S. Wolfram, MATHEMATICA: A program for doing Mathematics by Computer, Addison – Wesley, Reading, MA 1988.

[25] S. Karataglidis, B.A. Brown, K. Amos, and P.J. Dortmans, Phys. Rev. **C55** (1997) 2826.

[26] M. El-Azab Farid and G.R. Satchler, Nucl. Phys. **A438**(1985) 525.

[27] N. M. Clarke, Hi-optim 94.2 code (1994) University of Birmingham, England (unpublished).

[28] J.A. Tostevin and J.S. Alkhalili, Nucl. Phys.**A616** (1997) 418c.

[29] R.J. Glauber; Lectures on Theor. Phys. (Interscience, New York, 1959).

[30] Awad A. Ibraheem, Int. J. Mod. Phys**. E20** (2011) 721.

[31] Z. M. M. Mahmoud, Awad A. Ibraheem and M. El-Azab Farid, J. Phys. Scoc. Jap.**81** (2012) 124201.

[32] P. Shukla, Phys. ReV.**C**67 (2003) 054607.

[33] V.K. Lukyanov and E.V. Zemlyanaya, Int. J. Mod. Phys. **E10**,(2001) 169.